\documentclass[12pt,a4paper]{article}

\usepackage{epsfig}
\usepackage{amsmath,amsfonts,amssymb}
\usepackage{cite}
\usepackage{colortbl}
\usepackage{pifont}

\providecommand{\openone}{\leavevmode\hbox{\small1\kern-3.8pt\normalsize1}}

\newcommand{\afb}{A_\text{FB}}
\newcommand{\afbl}{A_\text{FB}^\ell}
\newcommand{\afbll}{A_\text{FB}^{\ell \ell}}
\newcommand{\ac}{A_C}
\newcommand{\acll}{A_C^{\ell \ell}}
\newcommand{\acl}{A_C^{\ell}}
\newcommand{\actl}{A_C^{t \ell}}
\newcommand{\cris}{c_{rs}}

\newcommand{\sla}[1]{/\!\!\!#1}

\begin{document}

\begin{center}
\begin{Large}
{\bf Single lepton charge asymmetries \\[2mm]
in $t \bar t$ and $t \bar t \gamma$ production at the LHC}
\end{Large}

\vspace{0.5cm}
J.~A.~Aguilar--Saavedra \\
\begin{small}
{ Departamento de F\'{\i}sica Te\'orica y del Cosmos, 
Universidad de Granada, \\ E-18071 Granada, Spain} \\ 

\end{small}
\end{center}

\begin{abstract}
We discuss lepton charge asymmetries in $t \bar t$ and $t \bar t \gamma$ production at the LHC, which can be  measured in the semileptonic decay channel $t \bar t \to W^+ b \, W^- \bar b \to \ell^+ \nu b \, q \bar q' \bar b$ (or the charge conjugate). Considering several variants of a new physics scenario with a light colour octet, it is seen that for $t \bar t$ these asymmetries may have a sensitivity competitive with the dilepton asymmetry already measured. For $t \bar t \gamma$ the new leptonic asymmetries, as well as the $t \bar t$ charge asymmetry, will reach their full potential with the high luminosity LHC upgrade. These asymmetries can pinpoint deviations at the $3\sigma-4\sigma$ level for new physics scenarios where the charge asymmetries already measured in $t \bar t$ production agree within $1\sigma$.  
\end{abstract}

\section{Introduction}

Charge asymmetries in $t \bar t$ production at hadron colliders~\cite{Aguilar-Saavedra:2014kpa} arouse wide interest after measurements by the CDF and D0 Collaborations of the forward-backward (FB) asymmetry $\afb$ in $t \bar t$ production at the Tevatron~\cite{Aaltonen:2011kc,Abazov:2011rq} showed some deviations from the existing standard model (SM) predictions~\cite{Kuhn:1998jr}. These anomalies  boosted the already extensive  program to measure the top quark properties, and in particular they provided a strong motivation for the measurement of a charge asymmetry $\ac$ in $t \bar t$ production at the Large Hadron Collider (LHC). 

Since then, the great expectations to find new physics effects in $t \bar t$ production have been cut down. The agreement of the $\ac$ measurements~\cite{Chatrchyan:2012cxa,Aad:2013cea} with the SM predictions~\cite{Kuhn:2011ri,Bernreuther:2012sx} brought some disappointment. Although the LHC $t \bar t$ charge asymmetry is a different observable (it is a forward-central rather than a FB asymmetry), it is somewhat correlated to $\afb$ in the SM and simple extensions~\cite{AguilarSaavedra:2011hz,AguilarSaavedra:2011ug}. It is possible to break this correlation~\cite{AguilarSaavedra:2012va,Drobnak:2012cz}, to achieve a large contribution to $\afb$ and negligible contribution to $\ac$, but doing that requires some tuning of parameters. A second blow for new physics expectations came with the updated Tevatron measurements using the full dataset~\cite{Aaltonen:2012it,Abazov:2014cca}, which showed a better agreement with the next-to-leading order (NLO) SM predictions, also refined with electroweak corrections~\cite{Hollik:2011ps,Kuhn:2011ri,Bernreuther:2012sx} and later improved to next-to-next-to-leading order (NNLO) accuracy~\cite{Czakon:2014xsa}.

\begin{figure}[t]
\begin{center}
\begin{tabular}{cc}
  \begin{tabular}{c}
  \includegraphics[width=7cm,clip=]{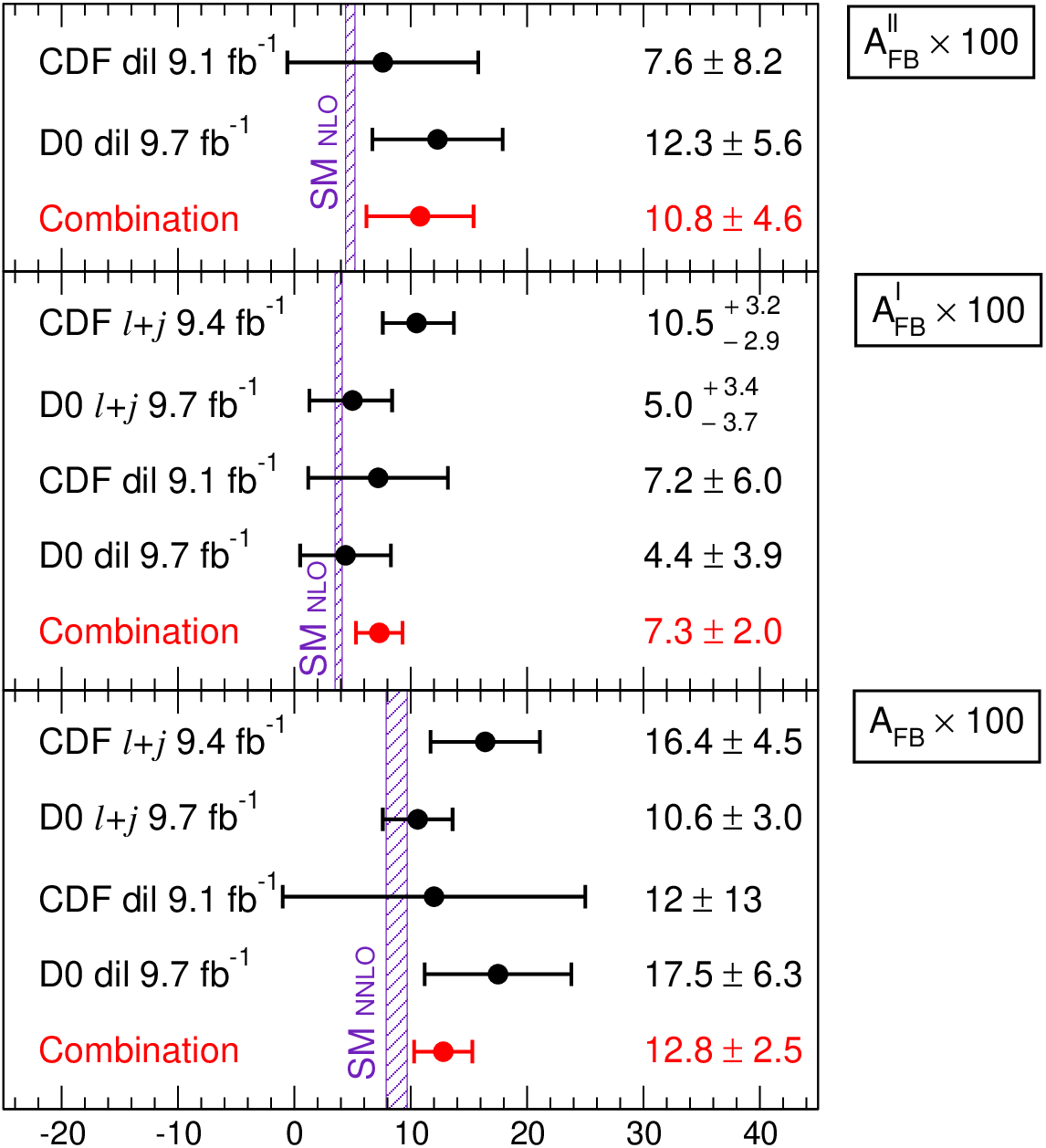}
  \end{tabular}
   &
  \begin{tabular}{c}
  \includegraphics[width=7cm,clip=]{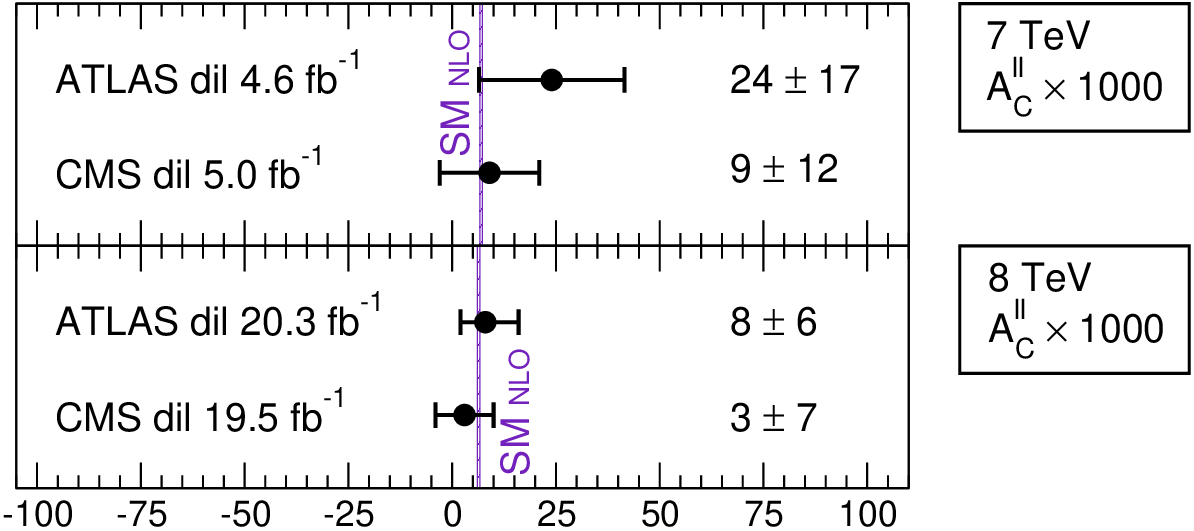} \\[5mm]
  \includegraphics[width=7cm,clip=]{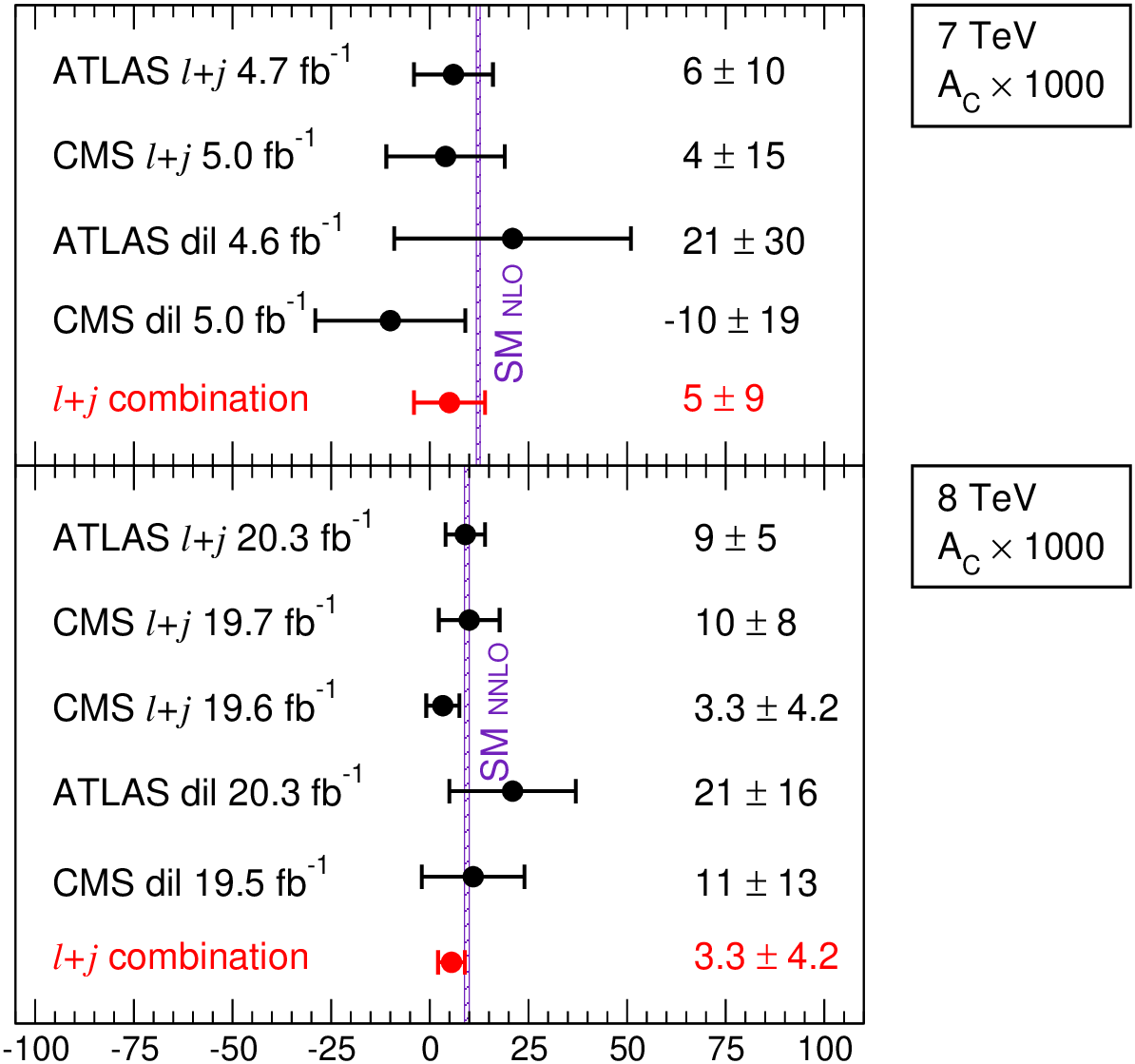}
  \end{tabular}
\end{tabular}  
\caption{Summary of FB and charge asymmetry measurements, compared to the SM predictions at NLO~\cite{Bernreuther:2012sx} and NNLO~\cite{Czakon:2014xsa,Czakon:2017lgo}. Official combinations are displayed in red.}
 \label{fig:summ1}
\end{center}
\end{figure}

At present, the combined measurements of Tevatron asymmetries~\cite{Aaltonen:2017efp} mostly agree with the SM predictions. There is a noticeable trend, with the four measurements of $\afb$ using the full data set~\cite{Aaltonen:2012it,Abazov:2014cca,Aaltonen:2016bqv,Abazov:2015fna}, as well as measurements of a related single lepton asymmetry $\afbl$~\cite{Aaltonen:2013vaf,Aaltonen:2014eva,Abazov:2014oea,Abazov:2013wxa} and a dilepton asymmetry $\afbll$~\cite{Aaltonen:2014eva,Abazov:2013wxa} always above the SM prediction, to varying degrees, see figure~\ref{fig:summ1} (left). LHC measurements at centre of mass (CM) energies of 7 TeV~\cite{Chatrchyan:2012cxa,Aad:2013cea,Chatrchyan:2014yta,Aad:2015jfa} and 8 TeV~\cite{Khachatryan:2015oga,Khachatryan:2015mna,Aad:2015noh,Khachatryan:2016ysn,Aad:2016ove}\footnote{Notice that the CMS Collaboration has performed two different measurements in the semi-leptonic channel with nearly the same dataset, using two different methods.} do not exhibit a clear trend (see the right panel) but there is some shift towards measurements {\it below} the SM prediction. In particular, among the naive combination of $\acll$ measurements at 7 TeV, the naive combination at 8 TeV, and the two $\ac$ official combinations at 7 and 8 TeV, three of these four measurements are found below the SM.

\begin{figure}[t]
\begin{center}
\includegraphics[height=7cm,clip=]{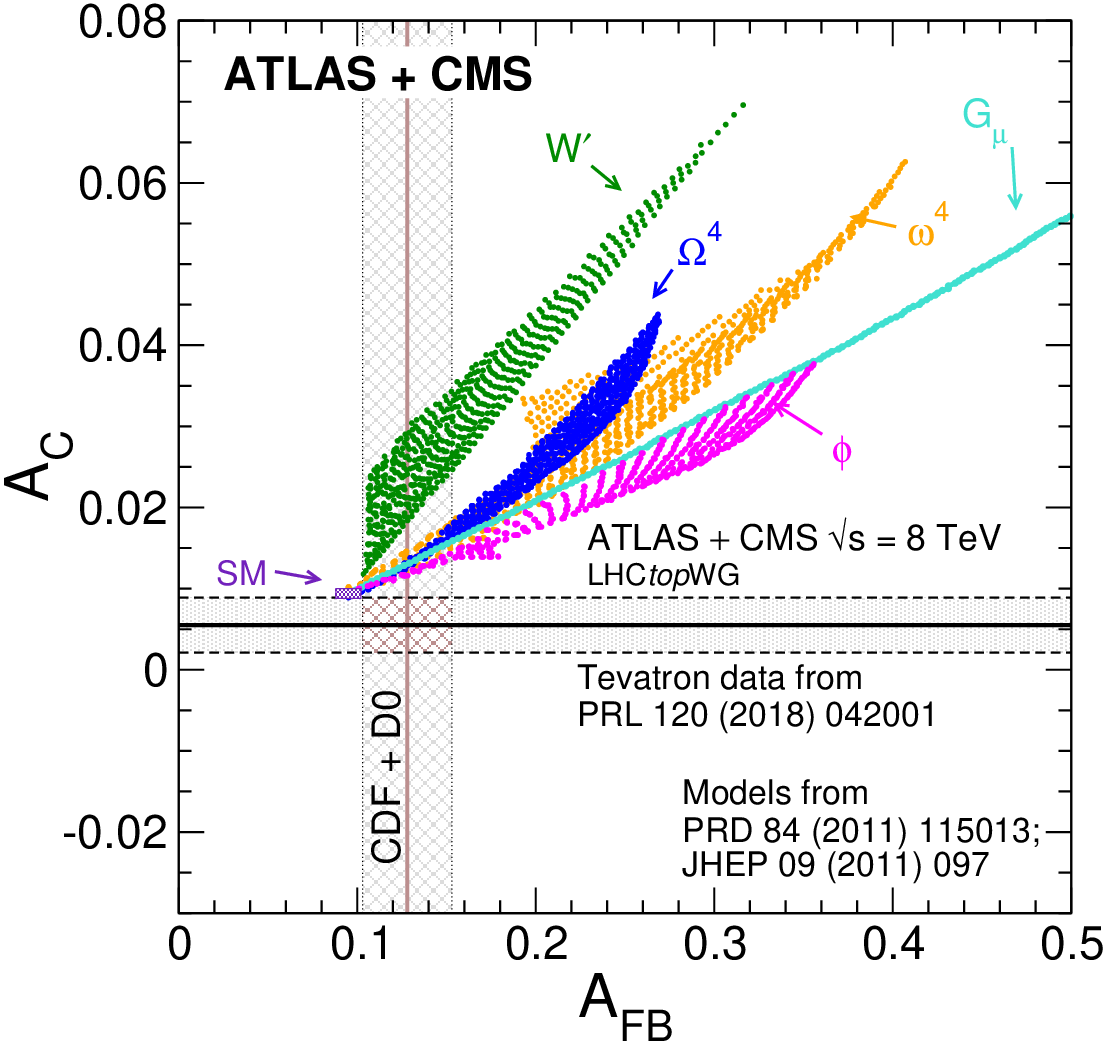}  
\caption{Measurements of $A_C$ and $\afb$, compared to the SM prediction and predictions from simple new physics models. From ref.~\cite{Sirunyan:2017lvd}.}
 \label{fig:summ2}
\end{center}
\end{figure}

The current status of the $t \bar t$ asymmetries, and the possible effect of simple new physics SM extensions, can be neatly summarised in figure~\ref{fig:summ2}, from ref.~\cite{Sirunyan:2017lvd}.
While the deviations are not significant, neither in the individual nor in the combined measurements, one cannot help but noticing the aforementioned trend, unlikely to arise from statistics alone. That might well be due to some mismodelling of $t \bar t$ production or, otherwise, some more contrived form of new physics, perhaps with different coupling to $u \bar u$ and $d \bar d$ since ---as it can be clearly seen from figure~\ref{fig:summ2}--- for simple new physics extensions the extra contributions to $\afb$ and $\ac$ are quite correlated and have the same sign.

The search for indirect new physics effects in the production and decay properties of the top quark continues at the LHC. Overall, the measurements agree with the SM, with some minor deviation in the top polarisation in the helicity axis, especially at 8 TeV, with the CMS measurement $P_z = -0.022 \pm 0.058$~\cite{Khachatryan:2016xws} and ATLAS results, $P_z = -0.044 \pm 0.038$ for top quarks and $P_z = -0.064 \pm 0.040$ for antiquarks~\cite{Aaboud:2016bit}, below the SM. Their naive combination, assuming CP conservation, is $P_z = -0.0477 \pm 0.0248$, consistent with zero at the $2\sigma$ level. In this context, one does not expect large deviations in any other $t \bar t$ observables as any new physics effect that could cause them would be severely constrained by existing measurements. But still, new observables are worth being explored as they could help identify if there might be something interesting behind the pattern of smallish deviations, or we are simply dealing with some mismodelling. It is then surprising that no single lepton charge asymmetry has yet been measured with LHC data. In this work we explore a laboratory frame asymmetry $\actl$~\cite{Carmona:2014gra} and a new $t \bar t$ rest frame asymmetry $\acl$, and their relation to the already measured $\ac$ and $\acll$.

The current Run 2 of the LHC will bring new opportunities with its higher statistics. The associated $t \bar t \gamma$ production will provide a measurement of a charge asymmetry (also denoted as $\ac$, for simplicity) that is independent from the one in $t \bar t$ production~\cite{Aguilar-Saavedra:2014vta}, because a hard photon emitted from the initial state couples differently to up and down quarks. Therefore, the ratio between  possible new physics contributions from $u \bar u$ and $d \bar d$ intial states is quite different
in $t \bar t$ and $t \bar t \gamma$. The charge asymmetry in $t \bar t W$ production~\cite{Maltoni:2014zpa} is not suppressed by the large (and symmetric) $gg \to t \bar t$ contribution, therefore it is more sensitive to new physics effects. In these processes, lepton asymmetries can be measured as well, providing extra independent probes on new physics effects. Charge asymmetries in $t \bar t j$ have also been proposed~\cite{Alte:2014toa,Berge:2013xsa}. In this work we focus on two single lepton asymmetries in $t \bar t \gamma$ production, $\actl$ and $\acl$, which are the analogue to the ones defined for $t \bar t$ production, and study their sensitivity. Their measurement in $t \bar t \gamma$ does not entail any extra combinatorial ambiguity, as might be the case for $t \bar t W$, and the only significant issue with respect to $t \bar t$ production is the need to suppress radiative top decays with an extra photon, which can be achieved by suitable kinematical cuts, as demonstrated in ref.~\cite{Aguilar-Saavedra:2014vta}. Because the measurements of these asymmetries are limited from statistics, we provide results for the high-luminosity Large Hadron Collider (HL-LHC) with 3 ab$^{-1}$, where the measurements will reach their full potential. An estimation of the sensitivity for the LHC with 300 fb$^{-1}$ can be obtained by a simple scaling.

\section{$t \bar t$ and leptonic charge asymmetries}
\label{sec:2}

The $t \bar t$ charge asymmetry measured by the ATLAS and CMS Collaborations is defined as
\begin{equation}
\ac = \frac{N(\Delta |y| >0) - N(\Delta |y| <0)}{N(\Delta |y| >0) + N(\Delta |y| <0)} \,,
\label{ec:ac}
\end{equation}
with $\Delta |y| = |y_t| - |y_{\bar t}|$ the difference between the moduli of the top quark and antiquark rapidities in the laboratory frame, and $N$ standing for the number of events. (As usual, the $\hat z$ axis in the laboratory frame is set in the beam direction.) The SM NLO prediction at 7 TeV is $A_C = 0.0123 \pm 0.0005$~\cite{Bernreuther:2012sx}, and the NNLO prediction at 8 TeV is $A_C = 0.0095^{+0.0005}_{- 0.0007}$~\cite{Czakon:2017lgo}.\footnote{This NNLO prediction uses the non-expanded denominator, as recently advocated. The prediction with expanded denominator is slightly larger, $A_C = 0.0097^{+0.0002}_{- 0.0003}$. When available, we use SM predictions with non-expanded denominators.} The combination of the ATLAS and CMS measurements in the semileptonic channel gives $A_C = 0.005 \pm 0.007\;\text{(stat)}\pm 0.006\;\text{(syst)}$ at 7 TeV and $A_C = 0.0055 \pm 0.0023\;\text{(stat)}\pm 0.0025\;\text{(syst)}$ at 8 TeV~\cite{Sirunyan:2017lvd}.

The dilepton charge asymmetry measured by the ATLAS and CMS Collaborations is
\begin{equation}
\acll = \frac{N(\Delta |y_\ell| >0) - N(\Delta |y_\ell| <0)}{N(\Delta |y_\ell| >0) + N(\Delta |y_\ell| <0)} \,,
\label{ec:acll}
\end{equation}
with $\Delta |y_\ell | = |y_{\ell^+}| - |y_{\ell^-}|$. The SM predictions at NLO are $\acll = 0.0070 \pm 0.0003$ at 7 TeV and $\acll = 0.0064 \pm 0.0003$ at 8 TeV~\cite{Bernreuther:2012sx}. Unlike the $t \bar t$ asymmetry, $\acll$ can only be measured in the dilepton decay channel (which is obvious from its definition). The naive combination of the results in figure~\ref{fig:summ1} gives $\acll = 0.013 \pm 0.010$ at 7 TeV and $\acll = 0.0057 \pm 0.0044$ at 8 TeV.

A laboratory frame single lepton asymmetry has been defined for the $t \bar t$ semileptonic decay channel in ref.~\cite{Carmona:2014gra},
\begin{equation}
\actl =  \frac{N(\Delta |y_{tl}| >0) - N(\Delta |y_{tl}| <0)}{N(\Delta |y_{tl}| >0) + N(\Delta |y_{tl}| <0)} \,,
\label{ec:actl}
\end{equation}
with $\Delta |y_{tl}| = q_\ell \left( |y_\ell| - |y_{t_h}| \right)$, $q_\ell = \pm 1$ the charge of the lepton and $t_h$ the top (anti-)quark decaying hadronically. A single lepton asymmetry in the $t \bar t$ rest frame can be defined as
\begin{equation}
\acl =  \frac{N(q_\ell \cris \cos \theta >0) - N(q_\ell \cris \cos \theta <0)}{N(q_\ell \cris \cos \theta >0) + N(q_\ell \cris \cos \theta <0)} \,,
\label{ec:acl}
\end{equation}
with $\theta$ the polar angle of the charged lepton in the $t \bar t$ rest frame, and
\begin{equation}
\cris = \text{sign}\; \vec r \cdot \vec s
\end{equation}
an additional $\pm 1$ factor introduced to break the symmetry in the $pp$ initial state, with $\vec r$ the three-momentum of the $t \bar t$ pair in the laboratory frame and $\vec s = \hat z$.\footnote{The $\hat z$ axis is set along the beam, in either (fixed) direction.} The advantage of the asymmetry (\ref{ec:actl}) is that its definition involves laboratory frame observables, therefore a reconstruction of the $t \bar t$ rest frame, with its associated uncertainties, is not necessary. The advantage of the asymmetry (\ref{ec:acl}) is its larger value, especially for some new physics benchmarks for which $\actl$ departs very little from the SM prediction. 

In the $t \bar t \gamma$ process the same definition for the charge asymmetry $\ac$ has been proposed~\cite{Aguilar-Saavedra:2014vta}. However, in order to achieve a good sensitivity to new physics contributions in the production, one has to design a suitable fiducial region where top radiative decays, i.e. $t \bar t$ production with a photon radiated from one of the top (anti-)quark decay products, is suppressed. This point has been verified for the semileptonic channel (see section~\ref{sec:4}). The charge asymmetry in $t \bar t \gamma$ arises at the tree level, and has a SM value $\ac = -0.033$ for the fiducial region used where radiative top decays are suppressed. Likewise, the same definition as for $t \bar t$ can be used for the single lepton asymmetries, but with the replacement of $\vec r$ by the three-momentum of $t \bar t \gamma$ in the definition of $\acl$.\footnote{Alternatively the three-momentum of $t \bar t$ can be used, leading to an asymmetry that is slightly smaller.} Their tree-level values are $\actl = -0.028$, $\acl = -0.022$. 

A dilepton asymmetry can also be defined for $t \bar t \gamma$ production, but its measurement faces two difficulties. First, the lack of sufficient statistics: even at the HL-LHC, the statistical uncertainty for a measurement in the dilepton channel will be around $\pm 0.006$. (For comparison, the total uncertainty of the $\ac$ measurement in $t \bar t$ at 8 TeV is $\pm 0.0042$.) Moreover, the definition of a suitable fiducial region where radiative top decays are suppressed requires the reconstruction of the momenta of the top quark and anti-quark, which is more difficult due to the two missing neutrinos, and will likely introduce additional uncertainties in the measurement, even if the charged lepton momenta are taken in the laboratory frame. For these two reasons, we do not expect this measurement to be competitive, and will omit a detailed analysis here.

\section{Predictions for $t \bar t$}
\label{sec:3}

We examine the potential of the new single lepton asymmetries by using as new physics benchmark a light colour octet below the $t \bar t$ threshold~\cite{AguilarSaavedra:2011ci} with mass $M= 250$ GeV and large width $\Gamma = 0.2 M$. This setup can evade constraints from other searches if its main decays are into new states~\cite{Tavares:2011zg,Gresham:2012kv} but, in any case, it is only used here for illustration of the potential deviations it produces in the several asymmetries. The relevant interaction Lagrangian is
\begin{equation}
\mathcal{L} = - \left[ \bar u \gamma^\mu {\textstyle \frac{\lambda^a}{2}} (g_V^u + \gamma_5 g_A^u) u 
+  \bar d \gamma^\mu {\textstyle \frac{\lambda^a}{2}} (g_V^d + \gamma_5 g_A^d) d 
+  \bar t \gamma^\mu {\textstyle \frac{\lambda^a}{2}} (g_V^t + \gamma_5 g_A^t) t 
\right] G_\mu^a \,,
\label{ec:lagr}
\end{equation}
in standard notation. The various constraints from $t \bar t$ observables on the parameter space of a light octet with this mass and width were analysed in full detail in ref.~\cite{Aguilar-Saavedra:2014nja}. Here we only include the most relevant constraints, namely
\begin{itemize}
\item the total cross section $\sigma$ at the Tevatron~\cite{Aaltonen:2013wca}, as well as the Tevatron asymmetries $\afb$, $\afbl$ and $\afbll$;
\item the top quark polarisation $P_z$ and the spin correlation coefficient $C$ in the helicity basis~\cite{Aaboud:2016bit,Khachatryan:2016xws} at the LHC with 8 TeV.
\end{itemize}
We require agreement within $2\sigma$ for these observables and study the allowed range for the new physics contributions to LHC asymmetries $\Delta \ac$, $\Delta \acll$, $\Delta \actl$ and $\Delta \acl$. (Since SM predictions are not available for the single lepton asymmetries, for consistency we always present our results in terms of the new physics contributions.) We collect the SM predictions~\cite{Czakon:2011xx,Czakon:2014xsa,Bernreuther:2012sx,Bernreuther:2015yna} and experimental measurements used in table~\ref{tab:input}.

\begin{table}[htb]
\begin{center}
\begin{tabular}{ccccccc}
& \multicolumn{2}{c}{Tevatron} &&& \multicolumn{2}{c}{LHC 8 TeV} \\
& prediction & measurement & \quad \quad & & prediction & measurement \\
$\sigma$ & $7.35^{+0.2}_{-0.24}$ pb & $7.60 \pm 0.41$ pb 
&& $P_z$ & 0 & $-0.0477 \pm 0.0248$ \\
$\afb$ & $0.095 \pm 0.007$ & $0.128 \pm 0.025$
&& $C$ & $0.318 \pm 0.003$ & $0.284 \pm 0.063$ \\
$\afbl$ & $0.038 \pm 0.003$ & $0.073 \pm 0.02$ \\
$\afbll$ & $0.048 \pm 0.004$ & $0.108 \pm 0.046$
\end{tabular}
\caption{Summary of SM predictions and experimental measurements used to constrain the parameter space of the light colour octet.}
\end{center}
\label{tab:input}
\end{table}

It has been shown~\cite{Aguilar-Saavedra:2014yea} that the Tevatron $t \bar t$ asymmetry $\afb$, the lepton asymmetries $\afbl$, $\afbll$, and the top polarisation,  are in general independent observables. In the benchmark model considered, their relation is somewhat dependent on the chirality of the colour octet coupling to the light quarks and the top quark. The same happens for LHC observables. For this reason, we will select nine representative benchmarks corresponding to light quarks with right-handed (R), left-handed (L) and axial (A) coupling to the octet (we take the same chirality for $u$ and $d$, for simplicity) and top quarks with R, L and A coupling to the octet. (For some of these chirality combinations a complicated model building may be required to obtain the corresponding couplings, as for example the left-handed $u$ and $d$ quarks have the same coupling to a colour octet $\text{SU}(2)_L$ singlet; however, this is not the goal of this phenomenological study.) For each of the chirality combinations we obtain the analytical dependence of the observables on the products $g^u g^t$ and $g^d g^t$ by calculating the observables for selected values of these products, using {\scshape Protos}~\cite{protos}. We then perform a fine scan over the parameter space to obtain our predictions. The couplings in the allowed regions are of order unity, for the mass $M= 250$ GeV used.

\begin{figure}[htb]
\begin{center}
\begin{tabular}{ccc}
\includegraphics[width=4.9cm,clip=]{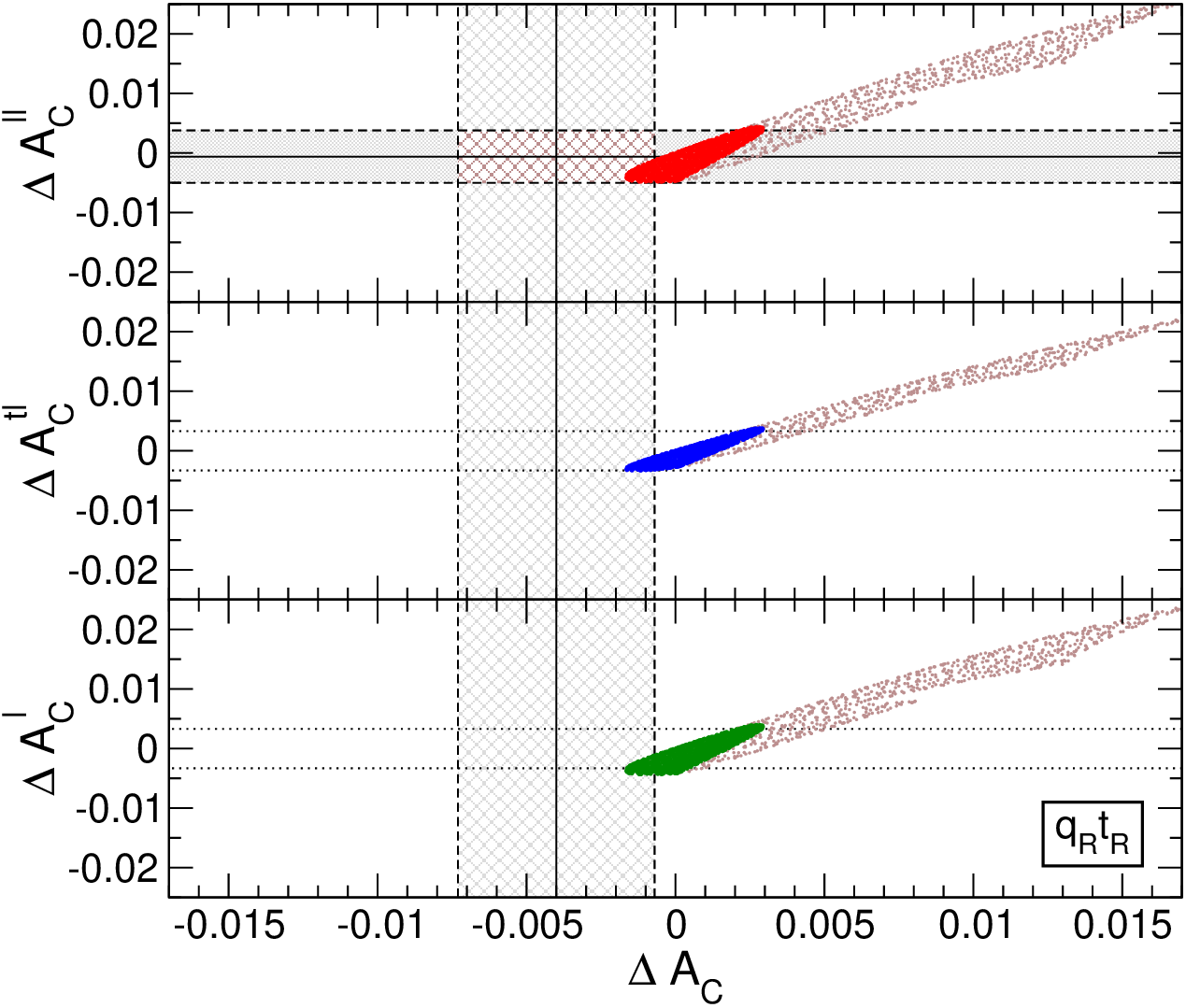} &
\includegraphics[width=4.9cm,clip=]{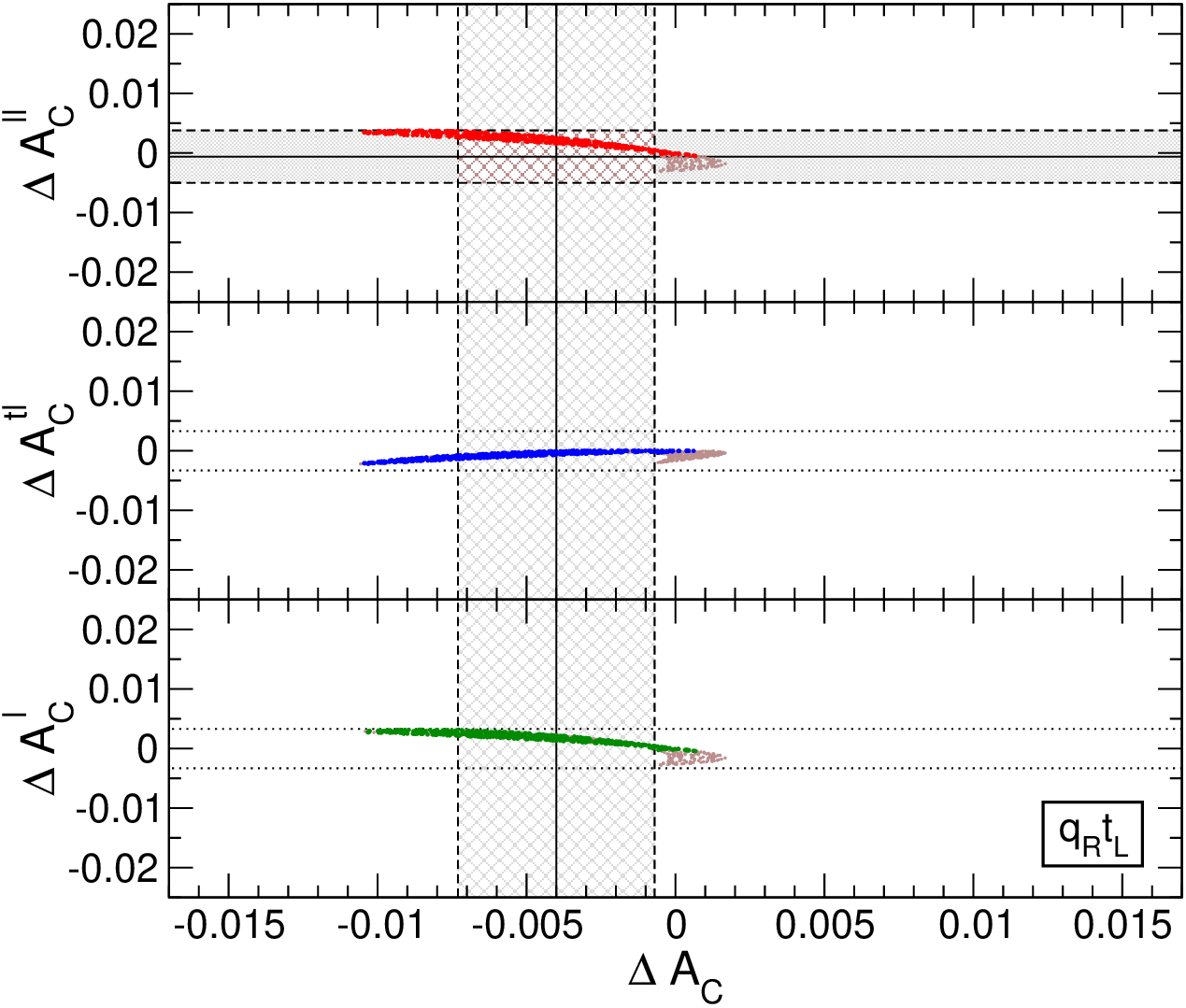} &
\includegraphics[width=4.9cm,clip=]{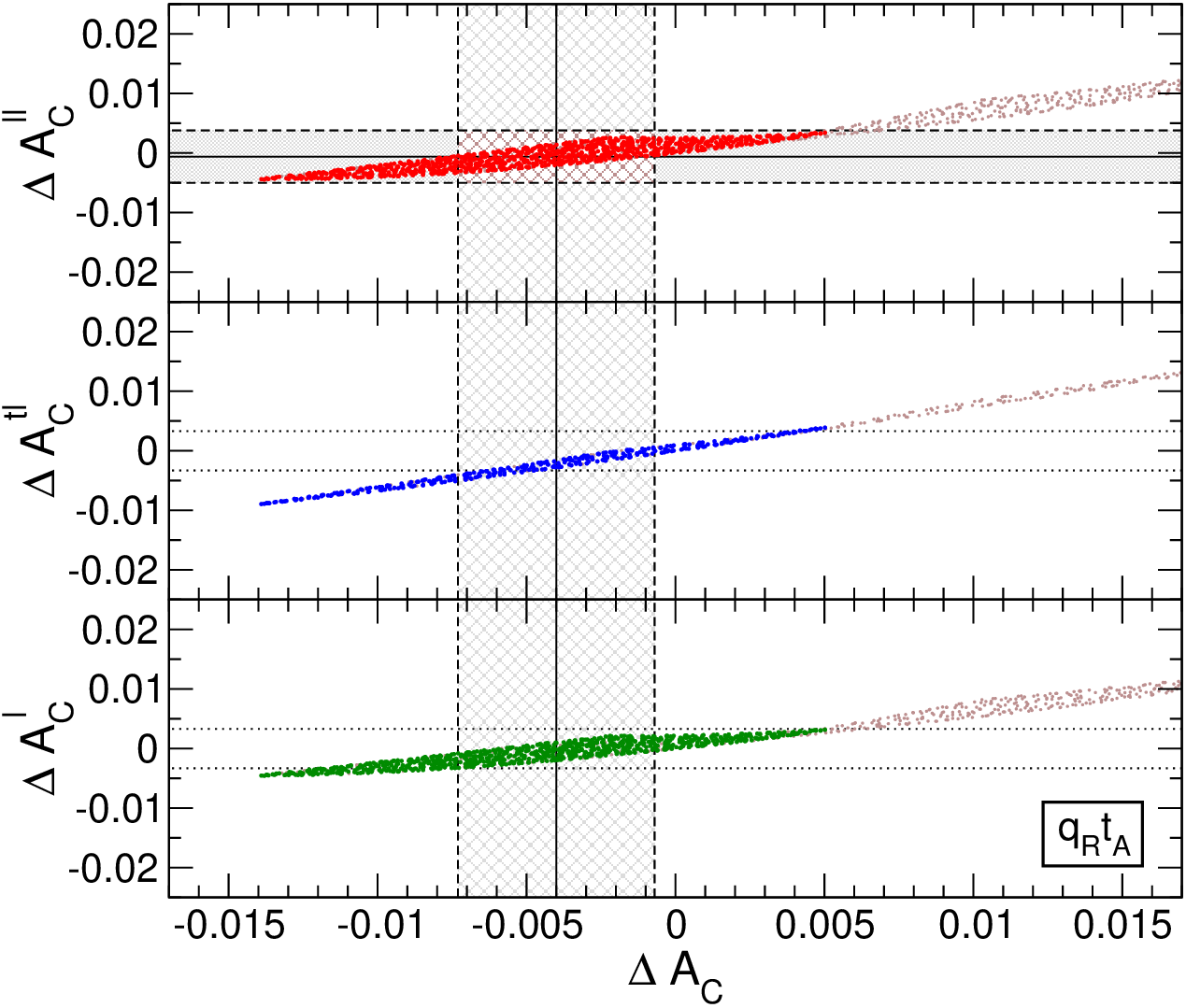} \\[2mm]
\includegraphics[width=4.9cm,clip=]{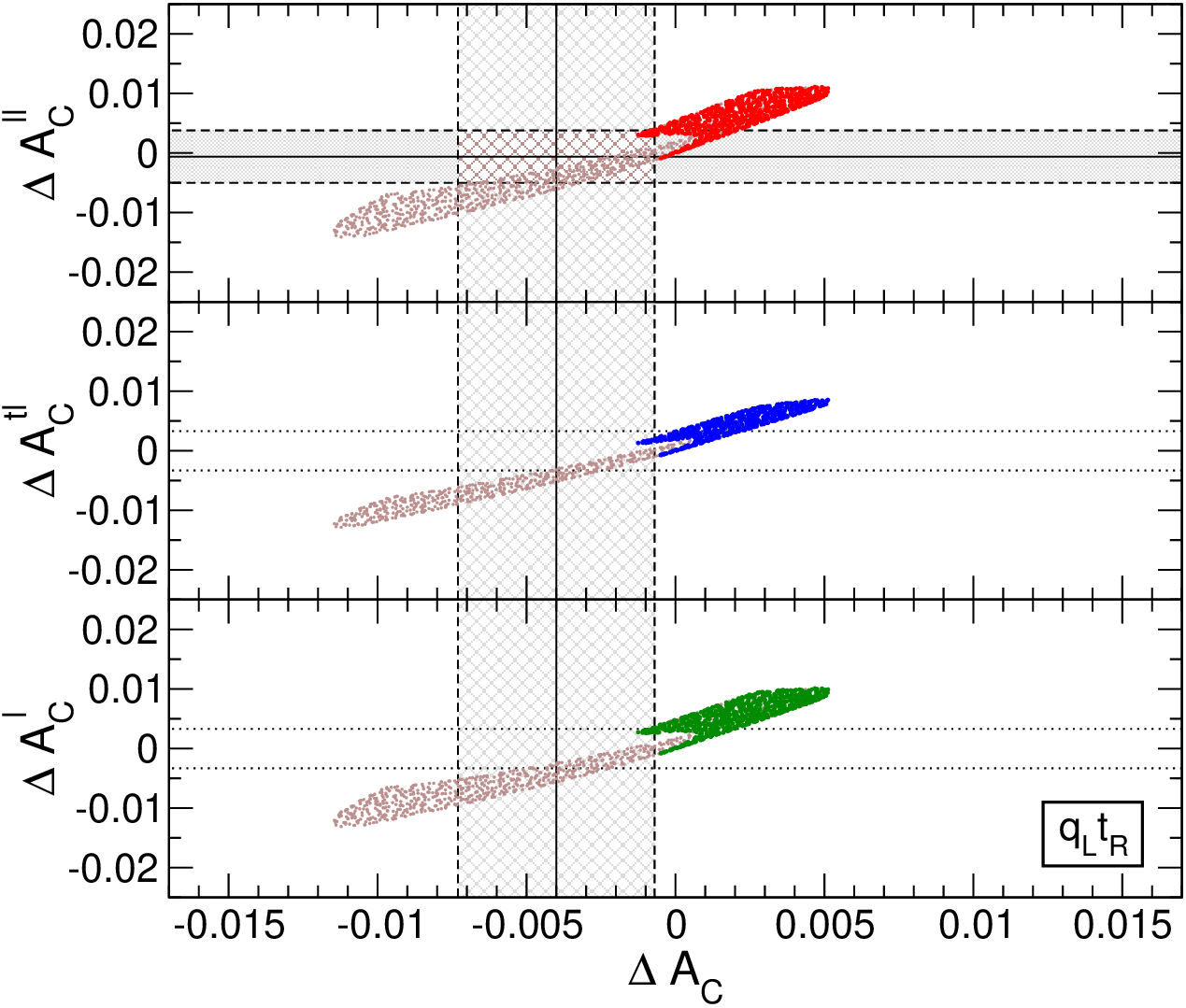} &
\includegraphics[width=4.9cm,clip=]{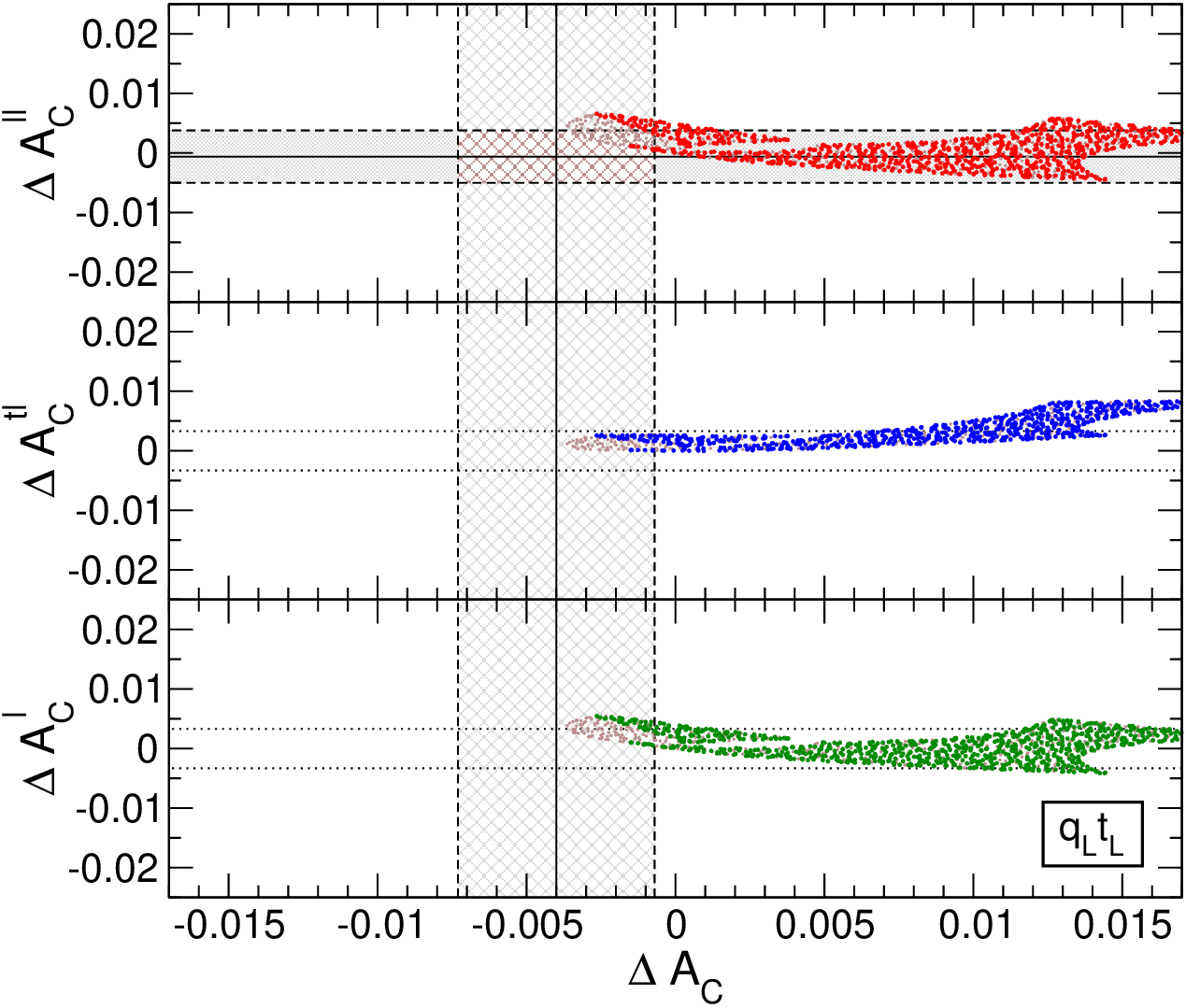} &
\includegraphics[width=4.9cm,clip=]{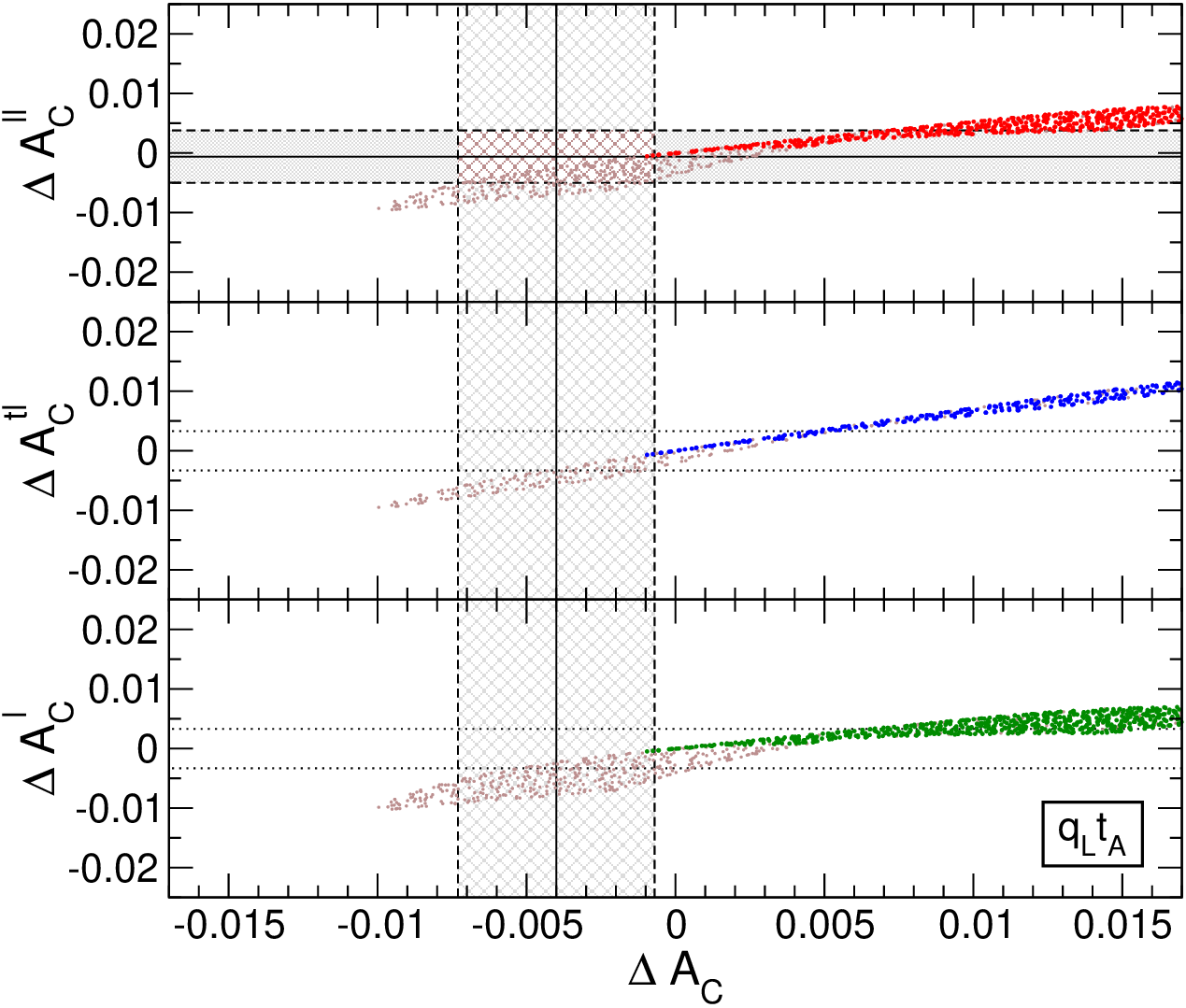} \\[2mm]
\includegraphics[width=4.9cm,clip=]{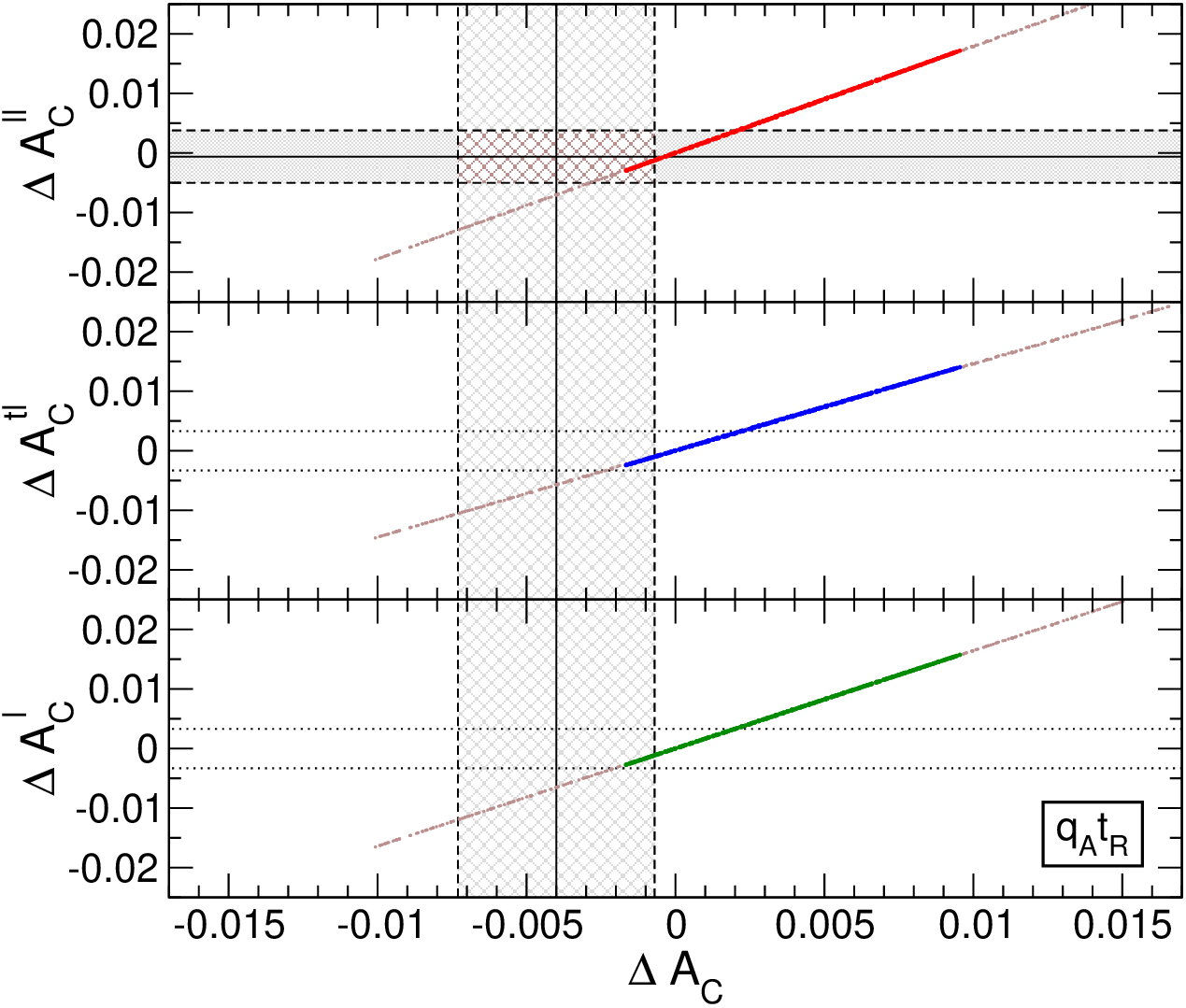} &
\includegraphics[width=4.9cm,clip=]{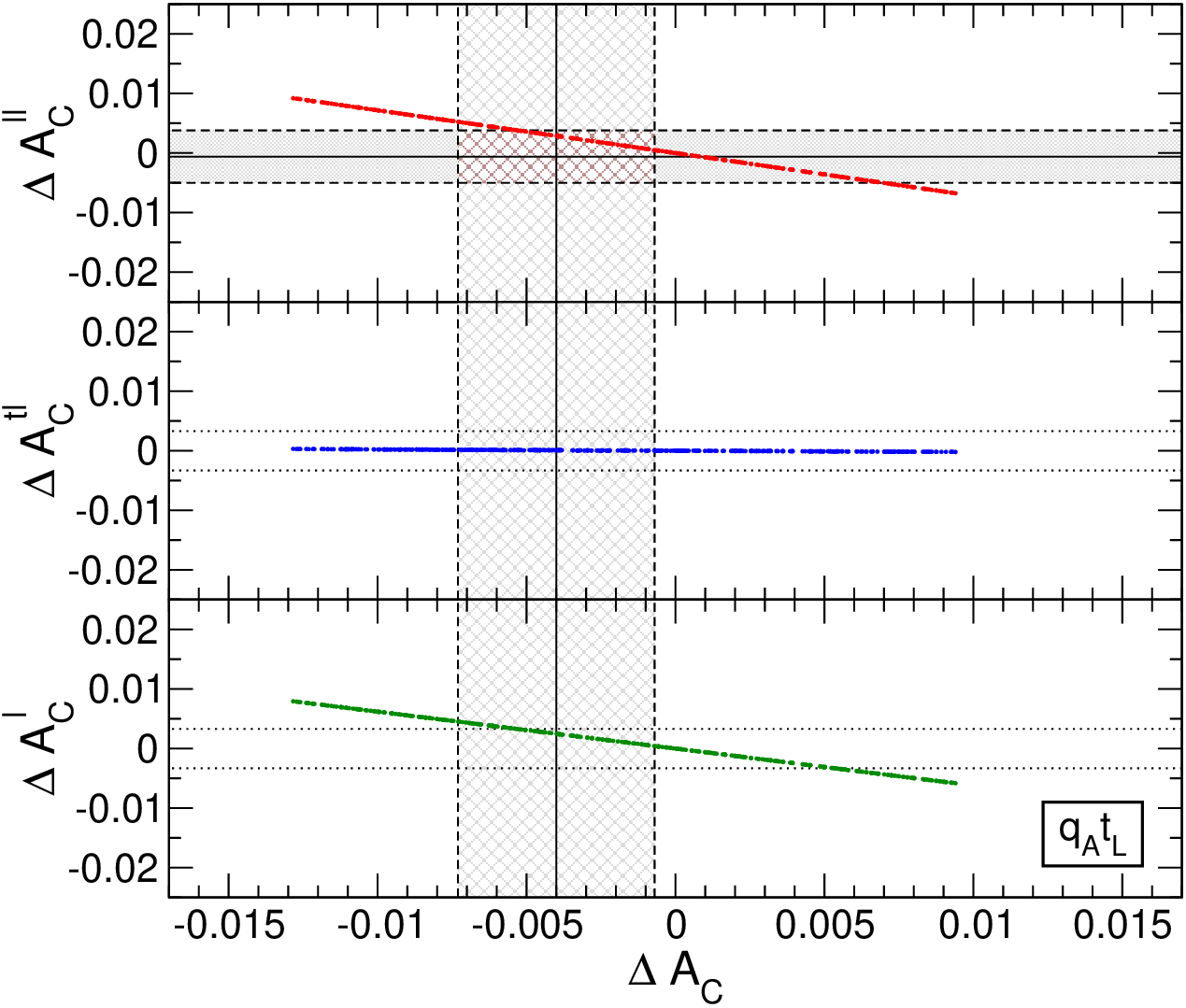} &
\includegraphics[width=4.9cm,clip=]{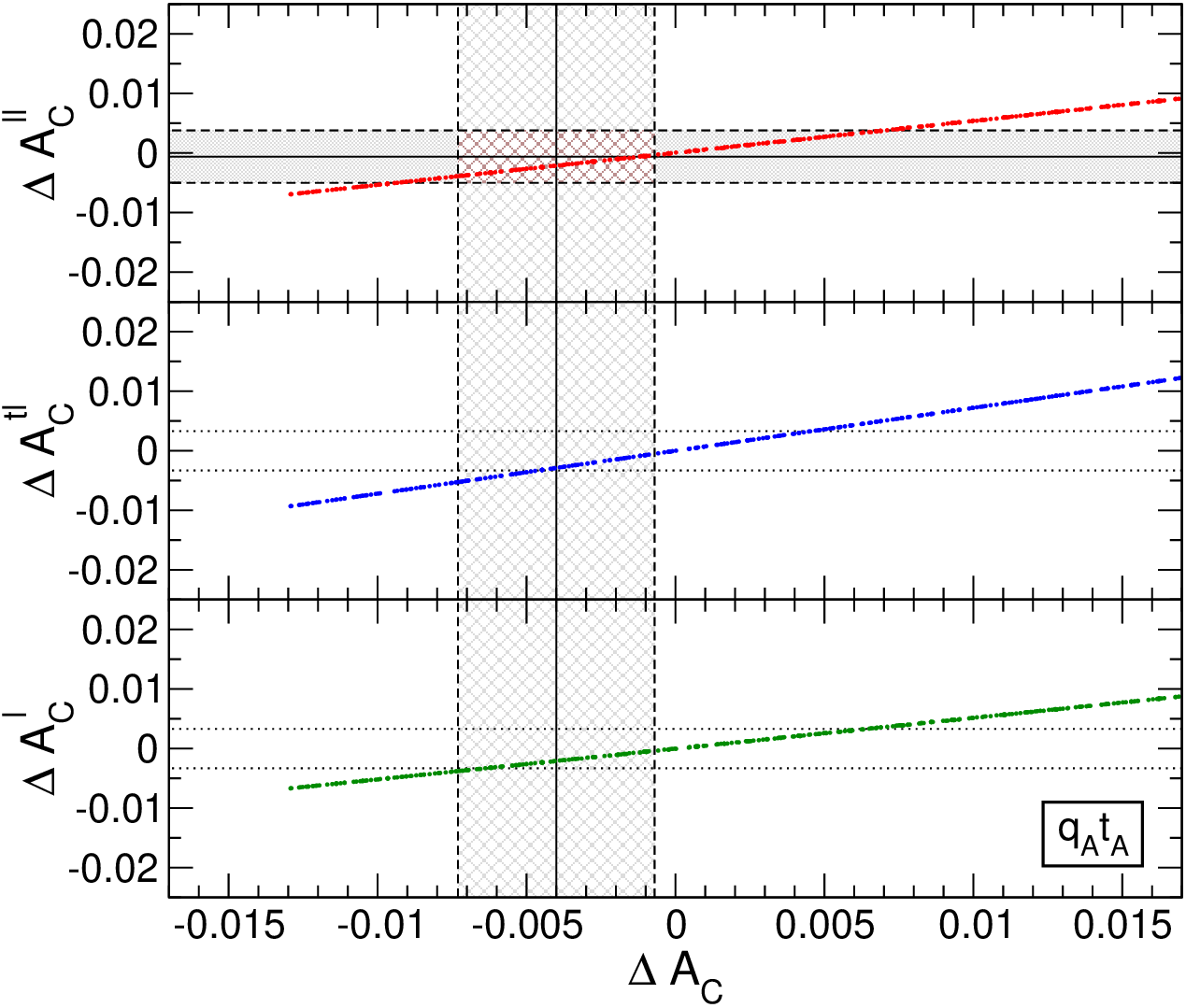} \\[2mm]
\end{tabular}  
\caption{Allowed range for the new physics contributions to $\acll$, $\actl$ and $\acl$, versus the new physics contributions to $\ac$, for the nine chirality benchmarks considered. The coloured regions include the constraint from $P_z$, while the gray regions do not.  The CM energy is 8 TeV.}
 \label{fig:tt}
\end{center}
\end{figure}

Our results for the lepton and $t \bar t$ asymmetries at the LHC are presented in figure~\ref{fig:tt}, for the nine chirality combinations, with a CM energy of 8 TeV. The vertical band represents the $\ac$ measurement and its $1\sigma$ uncertainty, with the SM contribution subtracted. In each panel, the top plot corresponds to $\Delta \acll$, with analogous definition, the middle plot to $\Delta \actl$ and the lower plot to $\Delta \acl$. The horizontal band for $\acll$ represents the naive combination of measurements in figure~\ref{fig:summ1} and its uncertainty, whereas for the single lepton asymmetries the band represents an estimation of the experimental uncertainty, taken equal to the current uncertainty in $\ac$. In the two latter cases, and for the rest of observables that have not yet been measured, the centre of the estimated uncertainty band is set at zero.
In the brown regions the requirement that $P_z$ lies within $2\sigma$ of its experimental value (which is already almost $2\sigma$ from the SM prediction) is dropped. The regions coloured in red, blue and green include that constraint. With this distinction, we can learn how the constraints would change if the agreement of $P_z$ with the SM prediction were better or, conversely, we can learn which type of new physics would provide a better overall agreement with all measurements. For example, for left-handed coupling to light quarks ($q_L$, middle panels) the agreement with the $P_z$ measurement results in a larger departure from the central values of $\ac$ and $\acll$. On the other hand, $q_R t_L$ and $q_R t_A$ contributions provide a better fit for all LHC measurements.

From the comparison of the allowed regions for $\acll$, $\actl$ and $\acl$ versus $\ac$ we observe that $\acll$ and $\acl$ are quite correlated or, in other words, they do not provide independent information. This fact does not render the measurement of $\acl$ useless, as (i) the precision in this single lepton asymmetry might be better; (ii) overconstraining the parameter space with independent measurements is quite useful when seeking for indirect hints of new physics. 
$\actl$ also follows the same pattern as $\acl$ and $\acll$ in most cases, except for $q_R t_L$ and $q_A t_L$ couplings, where it deviates very little from the SM prediction, while the other leptonic asymmetries do. This is quite expected, as this asymmetry is equivalent to $\ac$ but substituting the rapidity of either the top quark or antiquark by the rapidity of the corresponding charged lepton produced in the semileptonic decay.

\section{Predictions for $t \bar t \gamma$}
\label{sec:4}

Calculations for $t \bar t \gamma$ are computationally very demanding, so we restrict ourselves to a few selected benchmark points. Our interest here is to answer the question whether given the constraints imposed on the observables in table~\ref{tab:input}, plus good agreement with the measurements of $\ac$ and $\acll$, at the $1\sigma$ level, it would still be possible to obtain significant departures in $t \bar t \gamma$ asymmetries.

We perform our calculations with {\scshape MadGraph5}~\cite{Alwall:2014hca}, implementing the Lagrangian (\ref{ec:lagr}) in {\scshape Feynrules}~\cite{Alloul:2013bka} and interfaced to {\scshape MadGraph5} using the universal Feynrules output~\cite{Degrande:2011ua}. Following ref.~\cite{Aguilar-Saavedra:2014vta}, we use the following set of kinematical cuts to suppress radiative top decay:
\begin{itemize}
\item one lepton (electron or muon) with transverse momentum $p_T>20$~GeV and pseudorapidity $|\eta|<2.5$;
\item missing transverse momentum $\sla{p}_T>20$~GeV;
\item four quarks with $p_T>25$~GeV and $|\eta|<4.5$;
\item one photon with $p_T>20$~GeV and $|\eta|<2.5$;
\item lego-plot distance $\Delta R(\ell,\gamma)>1.0$, $\Delta R(\ell,j)>0.4$, $\Delta R(\gamma,q)>0.7$, $\Delta R(\gamma,b)>0.5$ $\Delta R(j,j)>0.4$, where $j$ denotes a light quark $q$ or a $b$ quark;
\item veto radiative $W$ decays: $m(jj\gamma)>90$~GeV, $m_T(\ell\gamma;\sla{p}_T)>90$~GeV,
where $m(jj\gamma)$ is the invariant mass of the $jj\gamma$ system and $m_T(\ell\gamma;\sla{p}_T)$ is the cluster transverse mass defined as
$$
m_T^2(\ell\gamma;\sla{p}_T)=\left(\sqrt{p_T^2(\ell\gamma)+m^2(\ell\gamma)}+ \sla{p}_T\right)^2-\left(\vec{p}_T(\ell\gamma)+\vec{\sla{p}}_T \right)^2 \,,
$$
with analogous definitions for particles other than the photon and the charged lepton;
\item veto radiative top decays: reject events satisfying either of the following conditions:
\begin{enumerate}
\item $m_T(b_{1,2}\ell\gamma;\sla{p}_T) < m_t + 20~{\rm GeV}$ and $m_t - 20~{\rm GeV} < m(b_{2,1}jj) < m_t + 20~{\rm GeV}$;
\item
$m_T(b_{1,2}\ell;\sla{p}_T) < m_t + 20~{\rm GeV}$ and $m_t - 20~{\rm GeV} < m(b_{2,1}jj\gamma) < m_t + 20~{\rm GeV}$,
\end{enumerate}
where $b_1, b_2 = b, \bar{b}$, and $b_1 \neq b_2$;
\item consistency with radiative top production: either 
\begin{enumerate}
\item $m_T(b_{1}\ell;\sla{p}_{\rm T}) < m_t + 20~{\rm GeV}$ and $m_t - 20~{\rm GeV} < m(b_{2}jj) < m_t + 20~{\rm GeV}$; or
\item $m_T(b_{2}\ell;\sla{p}_{\rm T}) < m_t + 20~{\rm GeV}$ and $m_t - 20~{\rm GeV} < m(b_{1}jj) < m_t + 20~{\rm GeV}$.
\end{enumerate}
\end{itemize}
The cross section for the semileptonic channel is of 81.5 fb at a CM energy of 14 TeV, after the cuts given above. With a luminosity of 3 ab$^{-1}$, and assuming a lepton triggering efficiency $\sim 70\%$, photon identification efficiency $\sim 85\%$, and $b$ tagging efficiency $\sim 70\%$, this results in a statistical uncertainty of $\pm 0.0027$. For the systematic uncertainty we can conservatively take a value twice larger than the one achieved in the 8 TeV measurement of $\ac$ in $t \bar t$, that is, $\pm 0.005$, resulting in a combined uncertainty of $\pm 0.0057$, which we take as estimation for all the asymmetry measurements in the semileptonic decay channel. The samples generated for each benchmark point (as well as the SM) have $5\times 10^5$ events, and the Monte Carlo uncertainty in the calculation of these asymmetries is $\pm 0.001$.

\begin{figure}[t]
\begin{center}
\includegraphics[height=7cm,clip=]{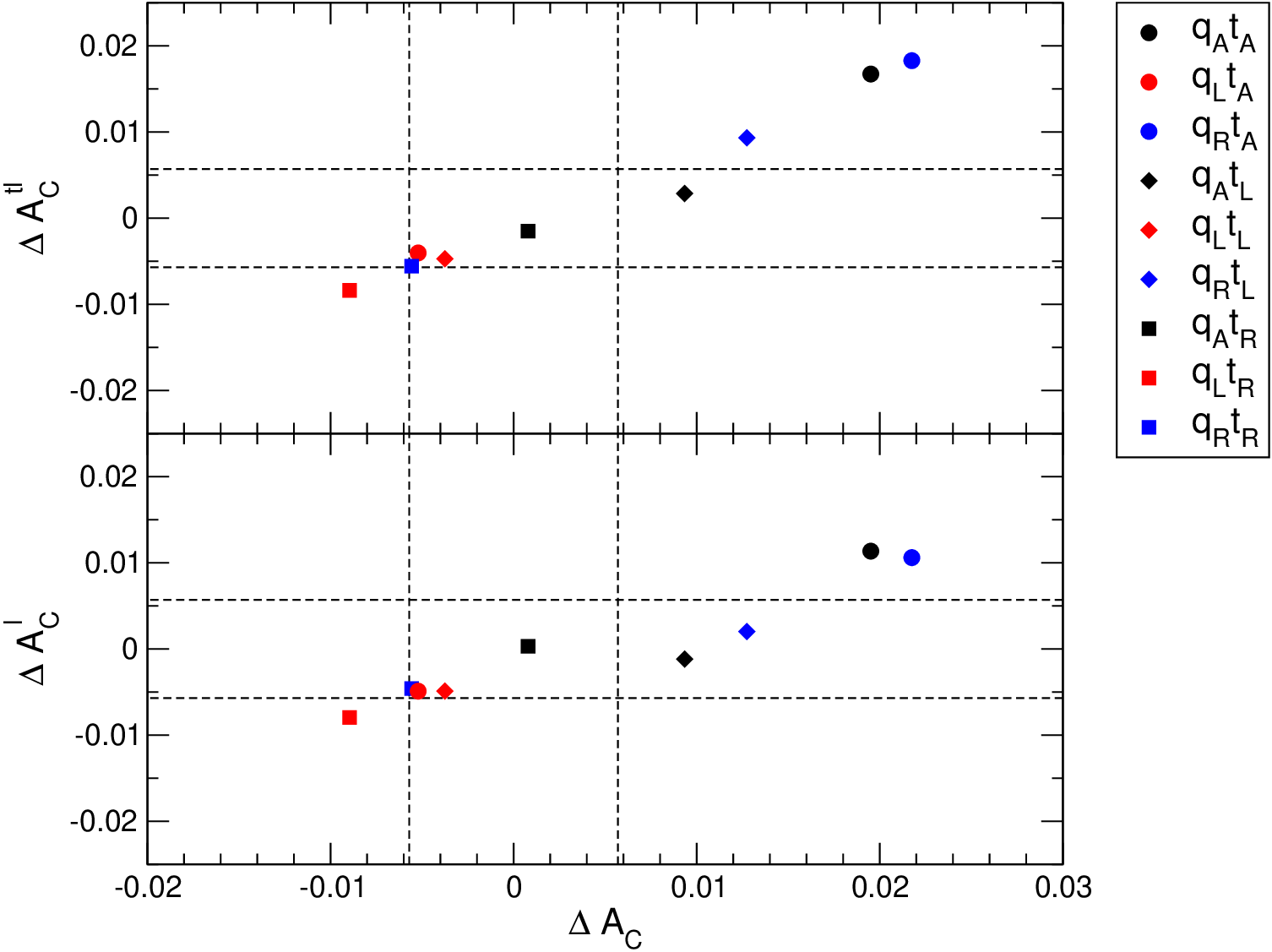}  
\caption{Allowed range for the new physics contributions to $\actl$ and $\acl$ in $t \bar t \gamma$, versus the new physics contributions to $\ac$ in $t \bar t \gamma$, for selected benchmark points corresponding to the the nine chirality combinations considered.  The CM energy is 14 TeV.}
 \label{fig:tta}
\end{center}
\end{figure}

Our results are presented in figure~\ref{fig:tta}, for $\Delta \actl$ versus $\Delta \ac$ (top) and $\Delta \acl$ versus $\Delta \ac$ (bottom). As before, the horizontal and vertical bands represent the estimated uncertainty of the asymmetry measurements. In agreement with earlier results~\cite{Aguilar-Saavedra:2014vta}, the departures in the $t \bar t$ asymmetry could be significant, up to the $3.8\sigma$ level. For the lepton asymmetries introduced in this work the deviations are also important, up to $3.2\sigma$ in $\actl$ and $2\sigma$ in $\acl$. These results highlight the complementarity and potential of the asymmetry measurements in $t \bar t \gamma$.

\section{Discussion}
\label{sec:5}

Independently of the possible anomalies in $t \bar t$ production at the Tevatron, $t \bar t \gamma$ production at the LHC offers a new window to test the top quark properties (see also ref.~\cite{Schulze:2016qas}).
The complementarity between $\ac$ in $t \bar t$ and $t \bar t \gamma$ to probe the presence of new physics contributions was already highlighted in ref.~\cite{Aguilar-Saavedra:2014vta}.
In this work we have revisited the potential of the $\ac$ measurement  in $t \bar t \gamma$, with an estimation for the HL-LHC, and obtained new estimations for the lepton asymmetries $\actl$ and $\acl$ in this process.

At the LHC run 2 with 13 TeV, these measurements are already very interesting. The $t \bar t \gamma$ cross section in the fiducial region considered in the previous section is slightly smaller than at 14 TeV, $68$ fb for the semileptonic decay channel. With the efficiency factors assumed, one expects a statistical uncertainty of $\pm 0.0095$ and a total uncertainty around $\pm 0.011$ in the measurements. Therefore, the potential deviations of the $t \bar t \gamma$ asymmetries in the benchmark points considered are up to $2\sigma$ in $\ac$ and $1.6\sigma$ in $\actl$. This is already remarkable, since for our estimations we have selected benchmark points that have $1\sigma$ agreement with the current very precise measurements of $\ac$ and $\acll$ in $t \bar t$ production. The full potential of the $t \bar t \gamma$ measurements will be reached at the HL-LHC, with potential deviations up to $3.8\sigma$ in $\ac$ and $3.2\sigma$, $2\sigma$ in $\actl$ and $\acl$, respectively.

Single lepton asymmetries in $t \bar t$ production are interesting as well. They can be measured at the current 13 TeV run, but for better comparison with existing measurements we have given results for 8 TeV. We have found that the predictions for $\acl$ and the dilepton asymmetry $\acll$ are quite correlated for the colour octet model considered, and less so for $\actl$, but their measurement is quite useful in any case, as (i) their experimental precision might be better; (ii) they are measured in a statistically independent sample, the semileptonic $t \bar t$ decay channel. Therefore, it might be worthwhile revisiting 8 TeV data to provide a measurement of these observables.

\section*{Acknowledgements}
I thank A. Juste for useful comments on the manuscript.
This work has been supported by MINECO Project  FPA 2013-47836-C3-2-P (including ERDF).



\begin{thebibliography}{99}

\bibitem{Aguilar-Saavedra:2014kpa}
  J.~A.~Aguilar-Saavedra, D.~Amidei, A.~Juste and M.~P\'erez-Victoria,
  {\it Asymmetries in top quark pair production at hadron colliders,}
  Rev.\ Mod.\ Phys.\  {\bf 87} (2015) 421
  [arXiv:1406.1798 [hep-ph]].

\bibitem{Aaltonen:2011kc}
  T.~Aaltonen {\it et al.} [CDF Collaboration],
  {\it Evidence for a Mass Dependent Forward-Backward Asymmetry in Top Quark Pair Production,}
  Phys.\ Rev.\ D {\bf 83} (2011) 112003
  [arXiv:1101.0034 [hep-ex]].

\bibitem{Abazov:2011rq}
  V.~M.~Abazov {\it et al.} [D0 Collaboration],
  {\it Forward-backward asymmetry in top quark-antiquark production,}
  Phys.\ Rev.\ D {\bf 84} (2011) 112005
  [arXiv:1107.4995 [hep-ex]].

\bibitem{Kuhn:1998jr}
  J.~H.~Kuhn and G.~Rodrigo,
  {\it Charge asymmetry in hadroproduction of heavy quarks,}
  Phys.\ Rev.\ Lett.\  {\bf 81} (1998) 49
  [hep-ph/9802268].

\bibitem{Chatrchyan:2012cxa}
  S.~Chatrchyan {\it et al.} [CMS Collaboration],
  {\it Inclusive and differential measurements of the $t \bar{t}$ charge asymmetry in proton-proton collisions at $\sqrt{s} =$ 7 TeV,}
  Phys.\ Lett.\ B {\bf 717} (2012) 129
  [arXiv:1207.0065 [hep-ex]].

\bibitem{Aad:2013cea}
  G.~Aad {\it et al.} [ATLAS Collaboration],
  {\it Measurement of the top quark pair production charge asymmetry in proton-proton collisions at $\sqrt{s}=$ 7 TeV using the ATLAS detector,}
  JHEP {\bf 1402} (2014) 107
  [arXiv:1311.6724 [hep-ex]].

\bibitem{Kuhn:2011ri}
  J.~H.~Kuhn and G.~Rodrigo,
  {\it Charge asymmetries of top quarks at hadron colliders revisited,}
  JHEP {\bf 1201} (2012) 063
  [arXiv:1109.6830 [hep-ph]].

\bibitem{Bernreuther:2012sx}
  W.~Bernreuther and Z.~G.~Si,
  {\it Top quark and leptonic charge asymmetries for the Tevatron and LHC,}
  Phys.\ Rev.\ D {\bf 86} (2012) 034026
  [arXiv:1205.6580 [hep-ph]].

\bibitem{AguilarSaavedra:2011hz}
  J.~A.~Aguilar-Saavedra and M.~P\'erez-Victoria,
  {\it Asymmetries in $t \bar{t}$ production: LHC versus Tevatron,}
  Phys.\ Rev.\ D {\bf 84} (2011) 115013
  [arXiv:1105.4606 [hep-ph]].

\bibitem{AguilarSaavedra:2011ug}
  J.~A.~Aguilar-Saavedra and M.~P\'erez-Victoria,
  {\it Simple models for the top asymmetry: Constraints and predictions,}
  JHEP {\bf 1109} (2011) 097
  [arXiv:1107.0841 [hep-ph]].

\bibitem{AguilarSaavedra:2012va}
  J.~A.~Aguilar-Saavedra and A.~Juste,
  {\it Collider-independent $t \bar t$ forward-backward asymmetries,}
  Phys.\ Rev.\ Lett.\  {\bf 109} (2012) 211804
  [arXiv:1205.1898 [hep-ph]].

\bibitem{Drobnak:2012cz}
  J.~Drobnak, J.~F.~Kamenik and J.~Zupan,
  {\it Flipping $t \bar t$ Asymmetries at the Tevatron and the LHC,}
  Phys.\ Rev.\ D {\bf 86} (2012) 054022
  [arXiv:1205.4721 [hep-ph]].

\bibitem{Aaltonen:2012it}
  T.~Aaltonen {\it et al.} [CDF Collaboration],
  {\it Measurement of the top quark forward-backward production asymmetry and its dependence on event kinematic properties,}
  Phys.\ Rev.\ D {\bf 87} (2013) no.9,  092002
  [arXiv:1211.1003 [hep-ex]].

\bibitem{Abazov:2014cca}
  V.~M.~Abazov {\it et al.} [D0 Collaboration],
  {\it Measurement of the forward-backward asymmetry in top quark-antiquark production in $p \bar p$ collisions using the lepton+jets channel,}
  Phys.\ Rev.\ D {\bf 90} (2014) 072011
  [arXiv:1405.0421 [hep-ex]].

\bibitem{Hollik:2011ps}
  W.~Hollik and D.~Pagani,
  {\it The electroweak contribution to the top quark forward-backward asymmetry at the Tevatron,}
  Phys.\ Rev.\ D {\bf 84} (2011) 093003
  [arXiv:1107.2606 [hep-ph]].

\bibitem{Czakon:2014xsa}
  M.~Czakon, P.~Fiedler and A.~Mitov,
  {\it Resolving the Tevatron Top Quark Forward-Backward Asymmetry Puzzle: Fully Differential Next-to-Next-to-Leading-Order Calculation,}
  Phys.\ Rev.\ Lett.\  {\bf 115} (2015) no.5,  052001
  [arXiv:1411.3007 [hep-ph]].

\bibitem{Czakon:2017lgo}
  M.~Czakon, D.~Heymes, A.~Mitov, D.~Pagani, I.~Tsinikos and M.~Zaro,
  {\it The top-quark charge asymmetry at the LHC and Tevatron through NNLO QCD and NLO EW,}
  arXiv:1711.03945 [hep-ph].

\bibitem{Aaltonen:2017efp}
  T.~A.~Aaltonen {\it et al.} [CDF and D0 Collaborations],
  {\it Combined Forward-Backward Asymmetry Measurements in Top-Antitop Quark Production at the Tevatron,}
  Phys.\ Rev.\ Lett.\  {\bf 120} (2018) no.4,  042001
  [arXiv:1709.04894 [hep-ex]].


\bibitem{Aaltonen:2016bqv}
  T.~A.~Aaltonen {\it et al.} [CDF Collaboration],
  {\it Measurement of the forward-backward asymmetry of top-quark and antiquark pairs using the full CDF Run II data set,}
  Phys.\ Rev.\ D {\bf 93} (2016) no.11,  112005
  [arXiv:1602.09015 [hep-ex]].

\bibitem{Abazov:2015fna}
  V.~M.~Abazov {\it et al.} [D0 Collaboration],
  {\it Simultaneous measurement of forward-backward asymmetry and top polarization in dilepton final states from $t\bar t$ production at the Tevatron,}
  Phys.\ Rev.\ D {\bf 92} (2015) 052007
  [arXiv:1507.05666 [hep-ex]].

\bibitem{Aaltonen:2013vaf}
  T.~A.~Aaltonen {\it et al.} [CDF Collaboration],
  {\it Measurement of the leptonic asymmetry in $t \bar t$ events produced in $p \bar p$ collisions at $\sqrt s=$ 1.96 TeV,}
  Phys.\ Rev.\ D {\bf 88} (2013) no.7,  072003
   Erratum: [Phys.\ Rev.\ D {\bf 94} (2016) no.9,  099901]
  [arXiv:1308.1120 [hep-ex]].

\bibitem{Aaltonen:2014eva}
  T.~A.~Aaltonen {\it et al.} [CDF Collaboration],
  {\it Measurement of the inclusive leptonic asymmetry in top-quark pairs that decay to two charged leptons at CDF,}
  Phys.\ Rev.\ Lett.\  {\bf 113} (2014) 042001
   Erratum: [Phys.\ Rev.\ Lett.\  {\bf 117} (2016) no.19,  199901]
  [arXiv:1404.3698 [hep-ex]].

\bibitem{Abazov:2014oea}
  V.~M.~Abazov {\it et al.} [D0 Collaboration],
  {\it Measurement of the forward-backward asymmetry in the distribution of leptons in $t\bar{t}$ events in the lepton$+$jets channel,}
  Phys.\ Rev.\ D {\bf 90} (2014) 072001
  [arXiv:1403.1294 [hep-ex]].

\bibitem{Abazov:2013wxa}
  V.~M.~Abazov {\it et al.} [D0 Collaboration],
  {\it Measurement of the asymmetry in angular distributions of leptons produced in dilepton $t\overline{t}$ final states in $p\overline{p}$ collisions at $\sqrt{s}=$ 1.96 TeV,}
  Phys.\ Rev.\ D {\bf 88} (2013) no.11,  112002
  [arXiv:1308.6690 [hep-ex]].



\bibitem{Chatrchyan:2014yta}
  S.~Chatrchyan {\it et al.} [CMS Collaboration],
  {\it Measurements of the $t\bar{t}$ charge asymmetry using the dilepton decay channel in pp collisions at $\sqrt{s} =$ 7 TeV,}
  JHEP {\bf 1404} (2014) 191
  [arXiv:1402.3803 [hep-ex]].

\bibitem{Aad:2015jfa}
  G.~Aad {\it et al.} [ATLAS Collaboration],
  {\it Measurement of the charge asymmetry in dileptonic decays of top quark pairs in $pp$ collisions at $\sqrt{s}=$ 7 TeV using the ATLAS detector,}
  JHEP {\bf 1505} (2015) 061
  [arXiv:1501.07383 [hep-ex]].



\bibitem{Khachatryan:2015oga}
  V.~Khachatryan {\it et al.} [CMS Collaboration],
  {\it Inclusive and differential measurements of the $t\overline{t}$ charge asymmetry in pp collisions at $\sqrt{s} =$ 8 TeV,}
  Phys.\ Lett.\ B {\bf 757} (2016) 154
  [arXiv:1507.03119 [hep-ex]].
  
\bibitem{Khachatryan:2015mna}
  V.~Khachatryan {\it et al.} [CMS Collaboration],
  {\it Measurement of the charge asymmetry in top quark pair production in pp collisions at $\sqrt s =$ 8 TeV using a template method,}
  Phys.\ Rev.\ D {\bf 93} (2016) no.3,  034014
  [arXiv:1508.03862 [hep-ex]].

\bibitem{Aad:2015noh}
  G.~Aad {\it et al.} [ATLAS Collaboration],
  {\it Measurement of the charge asymmetry in top-quark pair production in the lepton-plus-jets final state in pp collision data at $\sqrt{s}=$ 8 TeV with the ATLAS detector,}
  Eur.\ Phys.\ J.\ C {\bf 76} (2016) no.2,  87
  [arXiv:1509.02358 [hep-ex]].

\bibitem{Khachatryan:2016ysn}
  V.~Khachatryan {\it et al.} [CMS Collaboration],
  {\it Measurements of $t \bar t$ charge asymmetry using dilepton final states in pp collisions at $\sqrt s=$ 8 TeV,}
  Phys.\ Lett.\ B {\bf 760} (2016) 365
  [arXiv:1603.06221 [hep-ex]].

\bibitem{Aad:2016ove}
  G.~Aad {\it et al.} [ATLAS Collaboration],
  {\it Measurements of the charge asymmetry in top-quark pair production in the dilepton final state at $\sqrt{s}=$ 8 TeV with the ATLAS detector,}
  Phys.\ Rev.\ D {\bf 94} (2016) no.3,  032006
  [arXiv:1604.05538 [hep-ex]].




\bibitem{Sirunyan:2017lvd}
  M.~Aaboud {\it et al.} [ATLAS and CMS Collaborations],
  {\it Combination of inclusive and differential $\mathrm{t}\overline{\mathrm{t}}$ charge asymmetry measurements using ATLAS and CMS data at $\sqrt{s} =$7 and 8 TeV,}
  [arXiv:1709.05327 [hep-ex]].

\bibitem{Khachatryan:2016xws}
  V.~Khachatryan {\it et al.} [CMS Collaboration],
  {\it Measurements of $t \bar t$ spin correlations and top quark polarization using dilepton final states in pp collisions at $\sqrt s =$ 8 TeV,}
  Phys.\ Rev.\ D {\bf 93} (2016) no.5,  052007
  [arXiv:1601.01107 [hep-ex]].

\bibitem{Aaboud:2016bit}
  M.~Aaboud {\it et al.} [ATLAS Collaboration],
  {\it Measurements of top quark spin observables in $ t\overline{t} $ events using dilepton final states in $ \sqrt{s}=$ 8 TeV pp collisions with the ATLAS detector,}
  JHEP {\bf 1703} (2017) 113
  [arXiv:1612.07004 [hep-ex]].


\bibitem{Carmona:2014gra}
  A.~Carmona, M.~Chala, A.~Falkowski, S.~Khatibi, M.~Mohammadi Najafabadi, G.~Perez and J.~Santiago,
  {\it From Tevatron's top and lepton-based asymmetries to the LHC,}
  JHEP {\bf 1407} (2014) 005
  [arXiv:1401.2443 [hep-ph]].


\bibitem{Aguilar-Saavedra:2014vta}
  J.~A.~Aguilar-Saavedra, E.~\' Alvarez, A.~Juste and F.~Rubbo,
  {\em Shedding light on the $t \bar t$ asymmetry: the photon handle,}
  JHEP {\bf 1404} (2014) 188
  [arXiv:1402.3598 [hep-ph]].

\bibitem{Maltoni:2014zpa}
  F.~Maltoni, M.~L.~Mangano, I.~Tsinikos and M.~Zaro,
  {\it Top-quark charge asymmetry and polarization in $t\overline{t}W^±$ production at the LHC,}
  Phys.\ Lett.\ B {\bf 736} (2014) 252
  [arXiv:1406.3262 [hep-ph]].

\bibitem{Alte:2014toa}
  S.~Alte, S.~Berge and H.~Spiesberger,
  {\it Top quark charge asymmetry: searching for light axigluons in $ t\overline{t} $ + jet production at the LHC,}
  JHEP {\bf 1409} (2014) 084
  [arXiv:1406.0501 [hep-ph]].

\bibitem{Berge:2013xsa}
  S.~Berge and S.~Westhoff,
  {\it Top-Quark Charge Asymmetry Goes Forward: Two New Observables for Hadron Colliders,}
  JHEP {\bf 1307} (2013) 179
  [arXiv:1305.3272 [hep-ph]].


\bibitem{AguilarSaavedra:2011ci}
  J.~A.~Aguilar-Saavedra and M.~P\'erez-Victoria,
  {\it Shaping the top asymmetry,}
  Phys.\ Lett.\ B {\bf 705} (2011) 228
  [arXiv:1107.2120 [hep-ph]].

\bibitem{Tavares:2011zg}
  G.~Marques Tavares and M.~Schmaltz,
  {\it Explaining the $t \bar t$ asymmetry with a light axigluon,}
  Phys.\ Rev.\ D {\bf 84} (2011) 054008
  [arXiv:1107.0978 [hep-ph]].
  
\bibitem{Gresham:2012kv}
  M.~Gresham, J.~Shelton and K.~M.~Zurek,
  {\it Open windows for a light axigluon explanation of the top forward-backward asymmetry,}
  JHEP {\bf 1303} (2013) 008
  [arXiv:1212.1718 [hep-ph]].


\bibitem{Aguilar-Saavedra:2014nja}
  J.~A.~Aguilar-Saavedra,
  {\it Portrait of a colour octet,}
  JHEP {\bf 1408} (2014) 172
  [arXiv:1405.5826 [hep-ph]].

\bibitem{Aaltonen:2013wca}
  T.~A.~Aaltonen {\it et al.} [CDF and D0 Collaborations],
  {\it Combination of measurements of the top-quark pair production cross section from the Tevatron Collider,}
  Phys.\ Rev.\ D {\bf 89} (2014) no.7,  072001
  [arXiv:1309.7570 [hep-ex]].



\bibitem{Czakon:2011xx}
  M.~Czakon and A.~Mitov,
  {\it Top++: A Program for the Calculation of the Top-Pair Cross-Section at Hadron Colliders,}
  Comput.\ Phys.\ Commun.\  {\bf 185} (2014) 2930
  [arXiv:1112.5675 [hep-ph]].

\bibitem{Bernreuther:2015yna}
  W.~Bernreuther, D.~Heisler and Z.~G.~Si,
  {\it A set of top quark spin correlation and polarization observables for the LHC: Standard Model predictions and new physics contributions,}
  JHEP {\bf 1512} (2015) 026
  [arXiv:1508.05271 [hep-ph]].

\bibitem{Aguilar-Saavedra:2014yea}
  J.~A.~Aguilar-Saavedra,
  {\it Quantum coherence, top transverse polarisation and the Tevatron asymmetry $A^?_{FB}$,}
  Phys.\ Lett.\ B {\bf 736} (2014) 132
  [arXiv:1405.1412 [hep-ph]].

\bibitem{protos}
PROTOS, a PROgram for TOp Simulations. http://jaguilar.web.cern.ch/jaguilar/ protos/

\bibitem{Alwall:2014hca}
 J.~Alwall {\it et al.},
 {\it The automated computation of tree-level and next-to-leading order differential cross sections, and their matching to parton shower simulations,}
 JHEP {\bf 1407} (2014) 079
 [arXiv:1405.0301 [hep-ph]].

\bibitem{Alloul:2013bka}
  A.~Alloul, N.~D.~Christensen, C.~Degrande, C.~Duhr and B.~Fuks,
  {\it FeynRules  2.0 - A complete toolbox for tree-level phenomenology,}
  Comput.\ Phys.\ Commun.\  {\bf 185} (2014) 2250
  [arXiv:1310.1921 [hep-ph]].

\bibitem{Degrande:2011ua}
  C.~Degrande, C.~Duhr, B.~Fuks, D.~Grellscheid, O.~Mattelaer and T.~Reiter,
  {\it UFO - The Universal FeynRules Output,}
  Comput.\ Phys.\ Commun.\  {\bf 183} (2012) 1201
  [arXiv:1108.2040 [hep-ph]].

\bibitem{Schulze:2016qas}
  M.~Schulze and Y.~Soreq,
  {\it Pinning down electroweak dipole operators of the top quark,}
  Eur.\ Phys.\ J.\ C {\bf 76} (2016) no.8,  466
  [arXiv:1603.08911 [hep-ph]].

\end{thebibliography}
\end{document}